# Strategies for integrating uncertainty in iterative geostatistical seismic inversion

**Pedro Pereira†, Fernando Bordignon∗†, Leonardo Azevedo†, Ruben Nunes†**

**and Amílcar Soares†**

†*CERENA, Instituto Superior Técnico, University of*

*Lisbon, Portugal 1049-001.* ∗*Department of Informatics and Statistics, Federal University of Santa Catarina, Florianópolis, SC, Brazil, 88036-001*

Running head: **Integrating uncertainty into iterative inversion**

## ABSTRACT

Iterative geostatistical seismic inversion integrates seismic and well data to infer the spatial distribution of subsurface elastic properties. These methods provide limited assessment to the spatial uncertainty of the inverted elastic properties, overlooking alternative sources of uncertainty such as those associated with poor well-log data, upscaling and noise within the seismic data. We express uncertain well-log samples, due to bad logging reads and upscaling, in terms of local probability distribution functions (PDF's). Local PDF's are used as conditioning data to a stochastic sequential simulation algorithm, included as the model perturbation within the inversion. The problem of noisy seismic and narrow exploration of the model parameter space, particularly at early steps of the inversion, is tackled by the introduction of a cap on local correlation coefficients responsible for the generation of the new set of models during the next iteration. A single geostatistical framework is proposed and illustrated with its application to a real case study. When compared against a conventional iterative geostatistical seismic inversion, the integration of additional sources of uncertainty increases the match between real and inverted seismic traces and the variability within the ensemble of models inverted at the last iteration. The selection of the local PDF's plays a central role in the reliability of the inverted elastic models. Avoiding high local correlation coefficients at early stages of the inversion increases convergence in terms of global correlation between synthetic and real seismic reflection data at the end of the inversion.

# INTRODUCTION

Seismic inversion infers the spatial distribution of petro-elastic properties of the subsurface from recorded seismic reflection data. It is an ill-posed, nonlinear inverse problem with multiple solutions: many different elastic models can lead to the generation of highly similar synthetic responses that match considerably well the observed seismic reflection data (Tarantola, 2005). Consequently, any elastic model retrieved from seismic inversion is compromised by many sources of uncertainty such as geological uncertainties and data errors. Such uncertainties should be estimated and propagated consistently throughout and inversion procedure and taken into consideration during subsequent interpretation steps (e.g. see Bosch et al., 2010; Tompkins et al., 2011). Hence, a statistical framework is an adequate setting to address this problem.

Two of the main statistical-based seismic inversion methods are: Bayesian linearized seismic inversion (Buland and Omre, 2003; Grana and Della Rossa, 2010; Figueiredo et al., 2014; Grana et al. 2017) and iterative geostatistical seismic inversion procedures based on stochastic sequential simulation and co-simulation as the perturbation algorithm of the model parameter space (Bortoli et al., 1993; Soares et al., 2007; Nunes et al., 2012; Azevedo et al., 2013, 2015, 2018; Azevedo and Soares, 2017). Bayesian linearized inversions assume a linearized forward model and Gaussian, or multi-Gaussian, behavior representing both the expected variability of the elastic properties to be inferred from the inversion and the error term of the seismic data to be inverted. This approach results in the explicit solution of the posterior distribution to be Gaussian, or multi-Gaussian. Within this framework, the uncertainties related to the available data - both well-log and seismic - are expressed by the posterior mean and posterior model covariance matrix (Buland and Omre, 2003). In order to sample the model parameter space after the inversion procedure is finished, the posterior distribution can be sampled using Gibbs or Metropolis algorithms (Bosch et al., 2010).

On the other hand, iterative geostatistical seismic inversion avoids the need to assume any parametric distribution of the available data or the linearization of the forward model. Stochastic seismic inversion methods have the potential to integrate data with different levels of resolution, such as low vertical resolution seismic reflection with high vertical resolution well-log data and background trend models. Inverted elastic models with these methods exhibit high variability, reproducing the well-log data at their locations and allowing the assessment of the spatial uncertainty of the properties of interest (Doyen, 2007; Azevedo and Soares, 2017).

Stochastic sequential simulation and co-simulation algorithms, such as the Sequential Gaussian Simulation (Deutsch and Journel, 1998) and Direct Sequential Simulation (DSS; Soares, 2001), ensure that the available well-log data is exactly reproduced in all elastic models generated during the inversion. Also, all models show consistent spatial continuity patterns as revealed by a spatial covariance matrix (i.e., a variogram model within a two-point geostatistics framework). Each one of these models is commonly designated as a realization. Having been derived under the same assumptions regarding the number and locations of the available experimental data and the spatial continuity pattern these are considered equiprobable. Within this context, the existing experimental data represents hard data without any uncertainty. In these inversion methods, uncertain or unreliable experimental data, such as those due to measurement errors, are either omitted or retained as part of the conditioning data set without taking into consideration their intrinsic uncertainty.

Global stochastic inversion (GSI; Soares et al., 2007) is an iterative geostatistical seismic inversion method that uses DSS as the perturbation algorithm of the model parameter space and a global optimizer based on a cross-over genetic algorithm to search for a progressively better match between the observed and synthetic seismic data. The uncertainty is typically characterized by the ensemble of models produced at the final iteration, over which it is possible to compute statistics (e.g., the variance or the inter-quantile distance). Each realization of the elastic property of interest reproduces the available well-log data at the well locations and a variogram model that expresses the prior knowledge about the spatial continuity pattern of that property.

Honoring the well-log data in all elastic models generated during the inversion is, simultaneously, a positive and negative aspect of these kinds of inversion methods. Different sections of well-logs may have different levels of reliability, due for example to problems during well-log acquisition. In these cases, the well-log measurements should not be reproduced exactly in the inverted elastic models (Escobar et al. 2006, Williamson et al. 2007). However, the uncertainty associated with each measurement should be included and quantified within the inversion as they are useful for dictating the range and variability that can be expected in the inverted properties. In addition to potential uncertain log data, there is also the issue related to the upscaling of the acquired well-log data into the inversion grid. Iterative geostatistical seismic inversion methods require the upscaling of the high-frequency well-log data into the seismic scale. Usually a simple and deterministic upscaling technique such as Backus's average (Lindsay and Koughnet, 2001) of the original samples located within a coarser cell is used to represent the elastic properties as measured in the original well-log data. These procedures disregard uncertainty intrinsic to the upscaling process (e.g., the definition of the window length over which the average is calculated).

Another type of uncertainty often neglected in geostatistical inversion is the one related to the intrinsic uncertainty during the seismic processing and imaging steps (e.g., velocity modelling for migration). These methods use the trace-by-trace correlation coefficient between observed and synthetic seismic data to assess the goodness-of-fit of the simulated elastic model at a given iteration. The resulting local correlation coefficients control the generation of elastic traces during the next iteration. Thus, elastic traces that generate very high synthetic-to-real correlation coefficients are chosen as conditioning data for the next iteration, generating an ensemble of elastic models with very low variance at these trace locations. If high local correlation coefficients are reached at early stages of the inversion, they may lead to overfitting and a reduction in the exploration of the model parameter space.

We propose a geostatistical framework to account for these types of uncertainties in iterative geostatistical seismic inversion. The three application examples shown herein use the GSI, but the main concepts of the proposed approach are easily extrapolated to similar geostatistical inversion methods dealing with the elastic domain (e.g. Nunes et al., 2012; Azevedo et al., 2015; Azevedo et al., 2018). The first application example concerns uncertain well-log data where, for example, the borehole walls are unstable and well logging tools cannot obtain accurate readings. In the second example, we tackle the uncertainties related to the upscaling problem. Both issues are handled by modifying the conventional DSS algorithm, to account for local probability distribution functions (PDF's) rather than by using a single value at each log sample (Soares et al., 2017). The type and shape of these local distribution functions are directly related to the level of uncertainty associated with a given measurement for the property of interest. In the last application example, the issue of overfitting the observed seismic image, which essentially treats noise as though it were part of the signal, is managed by restricting the minimum variance allowed during the optimization. In this fashion, we can also deal with zones that have different noise levels throughout the seismic data grid.

## METHODOLOGY

We propose a GSI-based geostatistical seismic inversion to account for three different types of uncertainties that are frequently overlooked in iterative geostatistical inversion: 1) well-log measurement errors, which are frequently considered as valid experimental data; 2) uncertainty related to the upscaling procedure of the original well-log data into the coarser inversion grid; and 3) measurement errors present in the recorded seismic data that have not been mitigated or removed during processing and imaging.

We start by introducing the conventional global stochastic inversion method followed by the additional settings to integrate the proposed types of uncertainty.

### Global Stochastic Inversion

GSI is an iterative geostatistical seismic inversion method that uses direct sequential simulation and co-simulation (Soares, 2001) for model perturbation (Soares et al., 2007). In the first iteration, *Ns* realizations of acoustic impedance (Ip) are generated. The number of realizations, *Ns*, depends on the on the complexity of the subsurface geology and number of existing wells, but normally ranges between 16-32. The existing well-log data, upscaled into the inversion grid, are used as experimental data for the stochastic sequential simulation. A variogram model, which expresses the spatial continuity pattern of Ip in all three directions of space, is also imposed for the stochastic simulation. The variogram model is fitted to horizontal and vertical experimental variograms computed from the upscaled log samples and should reflect the expected spatial continuity pattern of Ip. While a background, or low-frequency model, is not explicit in this inversion method, the vertical and horizontal ranges of the variogram model imposed during the stochastic sequential simulation address this component implicitly. For each realization, the corresponding synthetics is computed by convolving the normal-incidence reflection coefficients with a wavelet. Then, at subsequent iterations, the best Ip model, generated at the end of the previous iteration, is used as a secondary variable along with the volume containing the local correlation coefficients for the stochastic sequential co-simulation of a new set of *Ns* Ip models. Depending on the local correlation coefficient at a given trace location, the new set of Ip models may exhibit larger or smaller variability within the ensemble. For locations with high correlation coefficients all models will be similar. The iterative procedure ensures the convergence of the synthetics towards the field data until a convergence criterion is reached, normally defined as the global correlation coefficient (Azevedo and Soares, 2017).

All the models generated by DSS during the GSI share the same characteristics: the reproduction of the global PDF based on upscaled Ip-log data; the reproduction of the values of the upscaled Ip-logs along the well paths; and a spatial continuity pattern as revealed by the imposed variogram model. For uncertainty assessment, we usually consider the ensemble of models generated during the last iteration, or those that generate high correlation between the seismic and synthetics, providing some important statistical elements such as mean and variance. This approach allows for assessment of the uncertainty related to the spatial distribution of the property of interest: areas of higher variance relate simultaneously to regions of the inversion grid that do not converge towards the observed seismic reflection data and areas far from the well locations.

To integrate additional types of uncertainty related to well-log data quality into the geostatistical inversion, we propose the use of direct sequential simulation with local PDF's (Soares et al., 2017). These types of uncertainties may be due to both measurement errors and upscaling. To control the variability of the models generated at a given iteration, avoiding trapping at local minima at early stages of the inversion, we propose an approach based on simulated annealing (Sen and Stoffa, 1991) to restrict the values of the local correlation coefficients, at the end of each iteration.

**Direct Sequential Simulation with local probability distribution functions**

Stochastic sequential simulation and co-simulation algorithms share three main characteristics: the simulated models honor the experimental data at their locations, they reproduce the original PDF as inferred from the experimental data, and they reproduce the variogram model fitted to the upscaled log samples that describe the spatial continuity pattern of the phenomena to be modeled (Goovaerts, 1997). In the presence of uncertain well-log data it is common practice to discard these data or decrease their weight within the simulation.

The DSS with local PDF's (Soares et al., 2017) is an extension of direct sequential simulation algorithm that accounts for the integration of uncertainty in the available experimental dataset. It assigns certain, $z(u_\alpha)$, and uncertain, $z(u_\beta)$, data in the simulation grid (Figure1a). This simulation algorithm assumes that the uncertainty associated with a given experimental sample located in $u_\beta$ can be modelled by a set of possible values, $z$, which occur according to a PDF $F(z(u_\beta))$. The local PDF's associated to those data samples can be derived for example by intelligent guessing or from secondary variables related to the property to be estimated. The type and shape of these local PDF's reflect the uncertainty attached to that measurement at a specific location (Figure 1b) Likewise, the experimental data without uncertainty $z(u_\alpha)$ and the corresponding global PDF for these data is defined as $F(z(u_\alpha))$ (Soares et al., 2017).

Before simulating the entire grid, this stochastic sequential simulation algorithm starts by randomly visiting the grid nodes located at $u_β$ (Figure 1c). At each sample location $u_β$, the simple Kriging estimate and kriging variance are calculated based on existing experimental data without uncertainty, $z(u_α)$, and previously simulated uncertain experimental data $z(u_β)$. These are used to build a local distribution $F(z(u_β))$ from where the next grid cell is drawn. After all uncertain locations, $u_α$, are visited, the remaining grid cells (Figure 1d) are simulated, through a random path, following direct sequential simulation using certain, $z(u_α)$, and previously simulated uncertain data, $z(u_β)$, as experimental data (Soares et al., 2017).

The models generated with this stochastic sequential simulation algorithm reproduce: 1) the spatial covariances as revealed by the imposed variogram model (Journel, 1994); 2) the local distributions at the locations of uncertain samples $u_β$; 3) the experimental data located at $u_α$ locations; and 4) the global distribution as inferred from the histogram of samples without uncertainty. A detailed description of this simulation algorithm is presented in Appendix A.

Including the DSS with local PDF's within an iterative geostatistical seismic inversion method enables the incorporation of uncertain well-log samples. Contrary to the normal DSS, these locations will exhibit different values for all the models generated from iteration-to-iteration. The values inverted at these locations will depend simultaneously on the a priori type of PDF assigned to each uncertain samples and the observed seismic reflection data. Contrary to conventional GSI, the resulting inverse solution comprises not only the uncertainty related to the non-unique nature of the seismic inverse problem but also the uncertainty associated with measurement errors of the experimental data. Therefore, at the end of the inversion we are able to retrieve a posterior PDF for these uncertain samples along the well path.

**Uncertain Well-log Data**

During drilling operations, sections of a borehole may collapse due to variations in the lithology (Figure 2). At these sections, the quality of the measured log data tends to decrease, being often unreliable, due to bad coupling of the logging tool with the borehole wall. This effect results in uncertainty of the measured property of interest. These poor log samples should ideally be identified as part of routine petrophysical editing and analysis and not be considered hard experimental data without any measurement error. In addition, there is also uncertainty from formations with thickness below the resolution limits of the logging tool. Both these uncertainties should be modeled and taken into consideration during seismic reservoir characterization.

To account for the uncertainty within collapsed zones we propose to remove the suspect well-log samples from the conditioning well-log dataset and to replace by local PDF's, reflecting the uncertainty associated with the log reading. These local PDF's can be inferred from: pseudo log values; prior knowledge based on expected values of the elastic properties of interest for a given geological setting; running alternative petrophysical models or results of an expeditious inversion at the well location using, for example, Bayesian linearized inversion (Figueiredo et al., 2014).

**Uncertainty in the Upscaling of Well-Log Data**

A common procedure for iterative geostatistical seismic inversion is to use an inversion grid with the same vertical resolution as the recorded seismic reflection data. For this reason, the measured high-resolution well-log data need to be upscaled into the inversion grid before being used as experimental data. The upscaling technique is performed by predicting the elastic rock properties at a lower frequency than the original log data (Bayuk et al., 2008). Information is always lost in the upscaling process, often resulting in a degradation of the fine scale accuracy of the predicted property of interest.

From a statistical point of view, upscaling characterizes a distribution of random variables, as interpreted from the measured well-log data, at a given cell of the coarser model by a single value. Resulting upscaled values should be representative of the high-resolution measurement, preserving its average elastic behavior. Figure 3 illustrates the contrast of using the simulation conditioning to a single Ip value at each log sample obtained by deterministic upscaling techniques (e.g., Backus's average) and the underlying PDF (i.e., the range of all possible values) of the high-resolution log within each coarser cell of the grid. The choice of the upscaling technique is important, because the upscaled samples along each well are reproduced in all models generated in geostatistical seismic inversion and can change considerably the original statistics as measured from the well-log data, resulting in possible inaccurate estimation of the elastic properties. Consequently, the upscaled elastic values must be consistent with both the observed seismic reflection data and the measured well-log.

Furthermore, it is important to recognize that the quality of the match between the seismic and synthetics at a well is not a sufficient criterion for judging the relative accuracy of the upscaling procedure. The reason for this ambiguity lies in the inherent non-uniqueness of the inversion result.

To account for the uncertainty during the upscaling, or thin bed effects (i.e., thin layers relative to the length of the tool), we propose that the upscaled well-log data are not used directly for the seismic inversion procedure; rather, they are replaced by the local PDF's for every cell of the inversion grid penetrated by the well paths. This avoids the problem of choosing a single value to characterize the cell distribution or to represent all the high-frequency measurements in a single average value. The posterior distribution at these locations is, therefore, conditioned by the a priori PDF's defined along the well path and the match between the resulting synthetic and observed traces at the well locations.

**Uncertainty in Recorded Seismic Reflection Data**

The convergence criteria for the GSI is defined by the global correlation coefficient (CC) between the best synthetic and the migrated full stack volume. The inversion converges because, at a given iteration, the elastic traces that generate the highest local CC are selected as secondary variable for the co-simulation of a new set of impedance models for the next iteration. The local CC are preponderant on the variability within the ensemble of models generated at a given iteration: Ip traces with high local CC values are preserved resulting in areas of low variability; while, Ip traces with low local CC values have low variability in the output.

To control the convergence rate of the inversion procedure we propose the use of a cap on the local collocated CC which increase at each iteration. At the last iteration, this limit is kept at 0.9 to avoid overfitting of the inverted elastic traces. This scheme is similar to the idea of the temperature in simulated annealing methods (Sen and Stoffa, 1995) where the sampling can accept worse solutions at the initial iterations, converging to a better final solution at the end of the optimization process.

# APPLICATION EXAMPLES

In this section, we show the results of three application examples in a real dataset and discuss the pros and cons of using the proposed stochastic sequential simulation as part of iterative geostatistical acoustic inversion when compared against the conventional GSI.

## Dataset Description

We use a real field data set composed of six vertical wells with a well log set containing caliper and Ip-logs, and a full stack seismic volume with 840×567×49 cells in *i*, *j* and *k* directions, respectively (Azevedo et al., 2013, Figure 4) with a bin size of 25 m by 25 m and a sampling rate of 4 ms. All six wells were used as conditioning data for all the inversion parameterization tested under the scope of this work. A wavelet extracted statistically from the available seismic reflection data was also made available and used as part of the inversion; however, uncertainty associated with the extraction procedure itself was not taken into consideration during the geostatistical inversion.

For illustrative purposes, the results of the three application examples in the next sections are presented locally using well W4 because it is the well that generates synthetics with the highest CC.

## Uncertain Well-log Data (Example 1)

The first application example shows the differences between the inverted Ip results for three distinct scenarios. In the first scenario, suspect well-log samples associated with collapsed borehole wall sections are excluded from the set of experimental data used in the stochastic sequential simulation. These locations are inverted, driven exclusively by the match between inverted and real seismic data using GSI. In the second and third scenarios, we assigned two types of local PDF's to these locations, and perturbed using DSS with local PDF's (Soares et al., 2017). In all three scenarios, six iterations and thirty-two realizations of Ip were generated per iteration using the same parameterization in terms of variogram model.

The first step was to exclude from the six wells the Ip-log samples where the caliper-log value is greater than 8.75in (0.22m; Figure 5a). This threshold value was chosen to be restrictive, maximizing the chances of the remaining well sections having smaller measurement errors due to irregular borehole walls.

Secondly, we prepared local distributions of Ip to replace the samples removed from the original well-logs. In the second scenario, suspect log samples are replaced by local uniform distributions centered at each log sample from the upscaled Ip-log and a spread equal to twice the standard deviation of all Ip-log upscaled values. This scenario considers large uncertainty at these sections. The third scenario comprises, local distributions inferred from the posterior distribution computed by performing a prior Bayesian linearized inversion (Figueiredo et al., 2014) at the closest trace to the well location (i.e., the local distributions are Gaussian with mean and variance inferred from Bayesian linearized seismic inversion).

Figure 5 compares the Ip traces inverted at the location of well W4 for the three scenarios described above. Two types of log data quality, or reliability, were interpreted along this well. One type consists log data where caliper samples have values above the pre-defined threshold, as indicated by the gray background shading in Figure 5: these log data are suspect. The other type consists of log data where caliper samples have values at or below the threshold, as indicated by the white background in Figure 5: these log data are considered reliable.

In conventional GSI (Figure 5b), when suspect log samples are removed from the set of conditioning data, the inverted values are inferred exclusively from the seismic reflection data at these locations. On the other two scenarios (Figure 5c and 5d), the samples considered suspect are not being inverted exclusively from the seismic reflection data, but they integrate a priori information as expressed by the shape and type of the local PDF's assigned at these locations. These reflect the range of a priori knowledge about the system we are modelling.

Regarding a global interpretation of Figure 5, for the zones generally associated with suspect log samples in the first scenario (Figure 5b), the range of Ip values obtained from the thirty-two realizations generated during the last iteration, including the P10 and P90 percentiles, is tight, and is not able to encompass the measured Ip-log. For the other two scenarios, where the uncertain samples were replaced by local PDF's (Figure 5c and 5d), the type of distribution does influence the variability of the inverted Ip traces and uncertainties. By assigning uniform PDF's, we were able to generate a final ensemble of models with high variability and Ip values closer to the measured Ip-log for the top section (Figure 5c). On the other hand, when considering information that is more precise, using the posterior distribution computed by linearized Bayesian inversion, the final ensemble of models produced lower variability (Figure 5d).

To facilitate further discussion, the well is split vertically into three main zones: the top zone, from the top of the logs to about 1,890 ms, generally includes suspect log data due to washouts; the middle zone, from about 1,890-1,940 ms and about 1,945-1,960 ms, generally includes reliable data; and the bottom zone, from about 1,960 ms to the bottom of the logs, again generally incudes suspect data. Taking into account the three scenarios simultaneously, in the middle zone of the well the inverted samples reproduce exactly the upscaled well-log data because these are used as experimental data and therefore all inverted logs overlap. A different behavior is verified in the top and bottom zones where the inverted samples are only partially consistent with the values from the original Ip-log. The inconsistency in behavior within both top and bottom zones might be due to two reasons: the inversion is converging to a local minima far from the true Ip values; or the original Ip-log is strongly contaminated with measurement errors and the true Ip values are different from the measured ones. Although the Ip-domain solutions achieved by the three scenarios represent alternative geological scenarios, to tackle this type of uncertainty, they are still not sufficient to determine the origin of the behavior inconsistency within these two zones. In order to obtain a more comprehensive explanation, we compared the local seismic responses between the real and synthetic seismic traces (Figure 6), where the seismic matches are good for all the scenarios within top and bottom zones. This consistency of inversion results in all the scenarios led to discard the reduced possibility of convergence to a local minima and to reinforce the idea of measurement errors as the origin of this type of uncertainty in top and bottom zones of the well. Finally, the guidance from seismic data and also the uncertainty integration based on the local PDF's of the last two scenarios helped us to understand the fidelity of prior removed well-log data: in fact log samples from 1800 ms to 1840 ms and from 1970 ms to 2000 ms are unreliable and log samples from 1840 ms to 1890 ms and from 1960 ms to 1970 ms are reliable. The latter case contradicts our initial expectation about the uncertainty in these zones, even in both situations the values are above the threshold of caliper-log, but it may be explained by the fact that the values of caliper-log are generally constant, from 1840 ms to 1890 ms, and just slightly above the threshold, from 1960 ms to 1970 ms.

**Uncertainty in the Upscaling of Well-Log Data (Example 2)**

The second application example integrates uncertainty related to the upscaling of the high-resolution well-log data into the inversion grid. In upscaling, there is always loss of information, with a common result that the extreme values present in the original well-log are not reproduced in the upscaled logs. From a geostatistical seismic inversion perspective, this may represent a drawback because the inverted elastic models will reproduce exactly the distribution of the upscaled well-logs (i.e., we might be under or overestimating the true Ip values).

Backus's averaging was used as the upscaling technique of Ip (Figure 7a). For the sake of simplicity, and for illustration purposes, we use Backus's averaging with a constant window length. In practice, the size of the window should change depending of the subsurface velocities. In the proposed framework, this can be accomplished by using depth-varying local PDF's.

To account for the loss of information caused by upscaling, local PDF's were assigned to the location of the reliable samples (samples outside the gray shading background in Figure 7). The same three scenarios of the previous example were considered: conventional GSI without local PDF's (Figure 7b); uniform PDF's centered at the upscaled well sample with a range twice the standard deviation of the high-resolution log (Figure 7c); and PDF's build from a Bayesian linearized inversion (Figure 7d).

In the GSI without local PDF's (Figure 7b), all the realizations honor the upscaled well log and consequently the average value of Ip for a given inversion cell grid. As previously discussed, in the sections corresponding to suspect log samples the results are driven exclusively by the mismatch between synthetic and real seismic.

Figure 7c illustrates the scenario when the upscaled samples are replaced by uniform PDF's for the upscaled samples. Figure 7d shows the results when using PDF's based on the a posteriori distribution of a linearized Bayesian inversion. Contrasting the first scenario with the last two scenarios in the interval between 1890 ms and 1960 ms, it is clear the existence of a range of values in each cell of the grid in the inversion results of latter cases, accounting for the variability of high-resolution. Furthermore, when compared to the homologous scenarios in Figure 5, it is visible the impact of integrating this type of uncertainty using the local PDF's, increasing the variability of values in top and bottom zones as well (even in the scenario using the solution from the Bayesian approach), due to a better exploration of model parameter space during the inversion.

**Uncertainty in Recorded Seismic Reflection Data (Example 3)**

The last application example compares the impact of using a cap in the local CC computed between synthetic and real traces at the end of each iteration. This aims consider uncertainty related to errors and assumptions during the seismic processing and imaging steps by avoiding overfitting. This additional criteria was implemented on top of the first application example where suspect log samples are removed from the conditioning dataset. The conventional GSI is used as the benchmark scenario for comparison.

Three distinct scenarios were tested using different capping criteria for the maximum local CC used during successive iterations (Table 1). The magnitude and sequence of caps were chosen based on the scenario of conventional GSI and are distinct between them in order to evaluate the impact on the variability of inversion results. Figure 8 shows the evolution of the global CC for the three scenarios, where it is visible that the application of chosen caps in the local CC distinctly impacts on the global convergence of the iterative procedures

The inverted traces at the location of well W4 (Figure 9) show that capping the local correlation coefficient results in more diversity within the final ensemble of models compared to the inversion scenarios of prior application examples. Acoustic traces in Figure 9a, which were generated without invoking any CC limit, have values smaller variability than those in Figures 9b-d, which were generated using CC caps described in Table 1. Use of high correlation coefficient at early iterations introduces a bias in the inversion procedure by limiting the exploration of the model parameter space. In addition, using consistently high correlation coefficient throughout the inversion (Figure 9d) reduces the variability in the inverted models as does the use of systematically lower correlation coefficients (Figures 9b and 9c). If during a particular inversion iteration, results get "trapped" in a local "valley" in the solution space, allowing lower caps on the correlation coefficients may help the solutions to essentially "jump" out of that valley.

The key parameter here may be the cap defined for the first iteration. Scenario 2 (Figure 9c) corresponds to the example with the lowest cap for the first iteration. Limiting the maximum trace-by-trace correlation coefficients to 50%, means that the models generated during the first iteration will have little influence in the co-simulation of iteration number two. However, it generates more variable models at the end of the inversion procedure. This ensures a larger exploration of the model parameter space (Figure 9c) and consequently reaches higher correlation coefficients when comparing against the other scenarios.

## DISCUSSION

The results of the application examples presented in the previous section were interpreted at a local scale (i.e., at the location of well W4). Here, we discuss the effect on the spatial distribution of Ip when including uncertain well samples. The results are illustrated through vertical well sections extracted from: the real and inverted seismic volumes (Figure 10); the average Ip model computed from the models simulated in the last iteration (Figure 11); the amplitude differences between real and the synthetics computed from the best-fit inverse Ip model (Figure 12); and the standard deviation of the ensemble of Ip models generated during the last iteration (Figure 13).

The synthetic seismic computed from the mean Ip model generated during the last iteration for the different application examples is similar in terms of reflection continuity (Figure 10). When defining local PDF's based on the high-resolution Ip-log (Figures 10e and 10f) we are also able to predict better the location and amplitude content of an amplitude anomaly highlighted by green ellipses.

The spatial continuity pattern of the inverse Ip models for all the application scenarios is globally similar (Figure 11). However, there are small-scale features that highlight the advantages of incorporating additional types of uncertainties. When compared with the conventional GSI, existing differences are mainly related with the ability to invert for extreme Ip values (Figure 11d and 11e). Extreme scenarios correspond to those situations with more potential impact during decision-making and should be considered during the geo-modeling workflow. These scenarios also correspond to those able to match the amplitude content of the reference seismic volume (Figure 10d and 10e). Another interesting feature that is barely perceptible in the conventional GSI scenario is the existence of a channel in the well path of W2 and highlighted by black ellipses (Figure 11). It is clear that using local PDF's allow to a better identification of channel location, especially in Figure 11d and Figure 11e. Regarding the zones of uncertain well-log data analyzed in Figure 5, they are highlighted in these vertical sections of Figure 11 by pink and red ellipses corresponding to the top and bottom zones of the well, respectively. In the former zone, the pink ellipses are illustrating that upscaled-log data and seismic reflection data are contradictory in the region adjacent to W4. In fact, the Ip values are a bit lower than those represented in the upscaled experimental data, which are also concordant to the inversion results of Figure 5. Red ellipses are pointing to an area where it is possible to interpret an abrupt discontinuity of a continuous layer close to this well. While none of the scenarios shown eliminates this effect completely, the use of local PDF's is able to mitigate it due to the integration of entire range values from high-resolution log, mainly in those scenarios when the variability is increased as well in the entire well (Figure 7c and Figure 7d; Figure 11d and 11e).This interpretation is supported by the computation of the amplitude differences between observed and synthetic seismic amplitudes (Figure 12). Consistently, the use of local PDF's for the entire well path (Figure 12d and 12e) ensure minimal residuals when comparing with the other application examples. This is probably caused by having more reliable Ip log samples at the well locations, which are going to condition the generation of Ip for locations in between the wells.

Figure 13 illustrates the standard deviation regarding all the scenarios showing the variability between the inverted models from last iteration. Brighter colors correspond to the regions with lower variability, contrasting to the darker colors which are related to the areas of higher variability and also higher uncertainty, generally way from the wells. An important aspect that needs to be taken into consideration is how the local PDF's impact the variability of the models generated at the last iteration of the inversion. In one hand, if selected distributions are too tight (Figures 13a to 13e), the variability of the ensemble is reduced with potential impact on the sampling of the plausible region of models. On the other hand, and taking into account the last application example (Figure 13f), the variability throughout the entire vertical section is much higher when compared with conventional GSI scenario and with the other four scenarios as well (Figure 13a to 13e). This result is related to the introduction of caps in the local CC, which were not applied to the other five examples. Finally, the other GSI scenarios using also the caps (local CC limits 1 and 3) are not illustrated here because they are very similar to the scenario of GSI with local CC limit 2 at a large-scale, and consequently they show a lower global impact between them regarding the variation of spatial continuity of Ip. By limiting the local correlation coefficients at the end of each iteration we are setting the inversion procedure in a more explorative nature when compared against the conventional GSI.

## CONCLUSION

In this work we tackled three types of uncertainties in geostatistical seismic inversion. We propose the use of a stochastic sequential simulation algorithm able incorporate local a priori information in the form of PDF's. These PDF's should reflect uncertainty associated with bad log reading or with the upscaling. Another critical aspect of these inversion methods is related to high trace-by-trace CC between synthetic and real seismic data at early stages of the inversion, which will stall the exploration of the model parameter space at a local minimum at the beginning of the inversion. This limits the explorative power of the inversion. The use of user-defined caps of maximum local CC allows these situations to be avoided. Results show that smaller caps at the first iterations generate higher correlation coefficients at the end of the inversion procedure.

This work intends to contribute on how uncertainties and measurements errors should be integrated in iterative geostatistical seismic inversion methods. Beside the uncertainties handled herein, an additional source of uncertainty that should be addressed is the one related with the variogram models used to describe the spatial continuity pattern of the properties of interest and imposed for the stochastic sequential simulation (Thore 2015).

## ACKNOWLEDGMENTS

The authors gratefully acknowledge the support of the CERENA (strategic project FCT-UID/ECI/04028/2013), Partex Oil & Gas for making this dataset available and Schlumberger for the donation of the academic licenses of Petrel®. FB would also like to thank CAPES for scholarship number 4041/15-1, and PETROBRAS.

## APPENDIX A

## DIRECT SEQUENTIAL SIMULATION WITH LOCAL PDF'S

Direct sequential simulation with local probability distribution functions (Soares et al., 2017) allows the generation of stochastic models integrating simultaneously experimental data without uncertainty (i.e., hard data), and uncertain experimental data. It is a two-step sequential simulation algorithm that can be summarized in the following sequence of steps.

1. Assign to each uncertain experimental data located in $u_\beta$, a local probability distribution function $F(z(u_\beta))$. This distribution function can be parametric or non-parametric and should reflect the uncertainty related to that measurement;

2. Generate a random path that visits sequentially all the locations of the uncertain experimental data $u_\beta$;

3. At each location $u_0$, along the random path defined in 2), compute the simple kriging estimate $(z(u_\beta)^*)$ and kriging variance $(\sigma_{SK}^2)$ based on the experimental data without uncertainty located within a neighborhood following:

$$z(u_0)^* = \sum_\alpha \lambda_\alpha(u_0)z(u_\alpha) + \sum_\beta \lambda_\beta(u_0)z(u_\beta) \tag{A-1}$$

$$\sigma_{SK}^2 = \frac{1}{n}\sum_{i=1}^{n}[z(u_i) - z(u_\mu)^*]^2 \tag{A-2}$$

4. Draw a value from the local probability distribution $F(z(u_\beta))$ within and interval centered on the simple kriging estimate $(z(u_0)^*)$ with a range equal to the kriging variance $(\sigma_{SK}^*)$;

5. Integrate the simulated value within the simulation grid as experimental hard-data;

6. Visit all the $u_\beta$ locations following the random path generated in 2 and repeat steps 3 to 5;

7. Define a random path within the simulation grid that visits all the cells without experimental data;

8. Visit all the locations defined in the random path generated in the previous step, and use direct sequential simulation algorithm to simulate values at these locations using the experimental hard-data and the previously simulated values at uncertain locations. At each realization a new set of uncertain data is generated.

This stochastic sequential simulation algorithm is able to generate realizations of the property of interest that reproduce the global distribution as inferred from the experimental hard data and the spatial continuity pattern as revealed by a variogram model. An ensemble of realizations of the property of interest also reproduces the local distributions at each location of the uncertain experimental data.

# LIST OF TABLES

# LIST OF FIGURES

respectively. a) High-resolution (in orange) and upscaled Ip-log data using Backus's averaging (in blue); Comparison between high-resolution log and inverted traces from: b) conventional GSI; c) GSI with local uniform distributions; and d) GSI with local distributions provided by linearized Bayesian inversion.

8      Evolution of the global CC between real and synthetic seismic volumes for the three scenarios shown in Table 1.

9      Well W4: gray background shadings represent suspect log samples; high-resolution Ip-log data is plotted in orange; 32 simulations in thin black; and P10, P50 and P90 percentiles in red, blue and green, respectively. Comparison between Ip traces inverted under the scenarios described in Table 1 for: a) conventional GSI without limitation criteria in the local CC; b) limit 1, c) limit 2 and d) limit 3.

10      Vertical seismic sections of a) real seismic data and the synthetics of: b) conventional GSI; c) GSI with local uniform distributions and d) GSI with local distributions provided by linearized Bayesian (Example 1); e) GSI with local uniform distributions and f) GSI with local distributions computed from linearized Bayesian inversion (Example 2) and g) GSI using as cap the local CC limit 2 (Table 1) (Example 3).

11      Vertical sections extracted from the average Ip model generated at the last iteration using: a) conventional GSI; b) GSI with local uniform distributions and c) GSI with local distributions provided by linearized Bayesian (Example 1); d) GSI with local uniform distributions and e) GSI with local distributions computed from linearized Bayesian inversion (Example 2) and f) GSI using as cap the local CC limit 2 (Table 1) (Example 3).

12      Vertical sections of amplitude residuals between real seismic data and synthetics of: a) conventional GSI; b) GSI with local uniform distributions and c) GSI with local distributions provided by linearized Bayesian (Example 1); d) GSI with local uniform distributions and e) GSI with local distributions computed from linearized Bayesian inversion (Example 2) and f) GSI using as cap the local CC limit 2 (Table 1) (Example 3).

13   Vertical sections of standard deviation illustrating the variability of the ensemble of 32 Ip models generated at the last iteration using: a) conventional GSI; b) GSI with local uniform distributions and c) GSI with local distributions provided by linearized Bayesian (Example 1); d) GSI with local uniform distributions and e) GSI with local distributions computed from linearized Bayesian inversion (Example 2) and f) GSI using as cap the local CC limit 2 (Table 1) (Example 3).

**Pereira et al** –

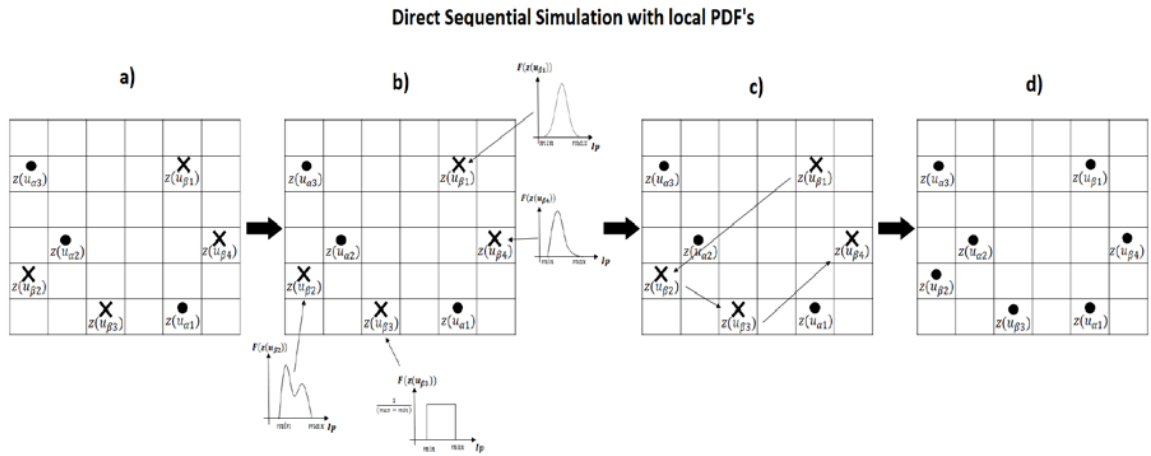

**Figure 1:** Schematic representation of DSS with local PDF's: a) assign uncertain ($z(u_\beta)$) and certain ($z(u_\alpha)$) experimental data to the simulation grid; b) assign local PDF's to the location of uncertain experimental data ($z(u_\beta)$); c) simulate all locations of uncertain data following a random path that visit all location of uncertain samples; and d) simulate remaining grid locations using $z(u_\alpha)$ and $z(u_\beta)$ as experimental data and following a pre-defined random path.

**Pereira et al**

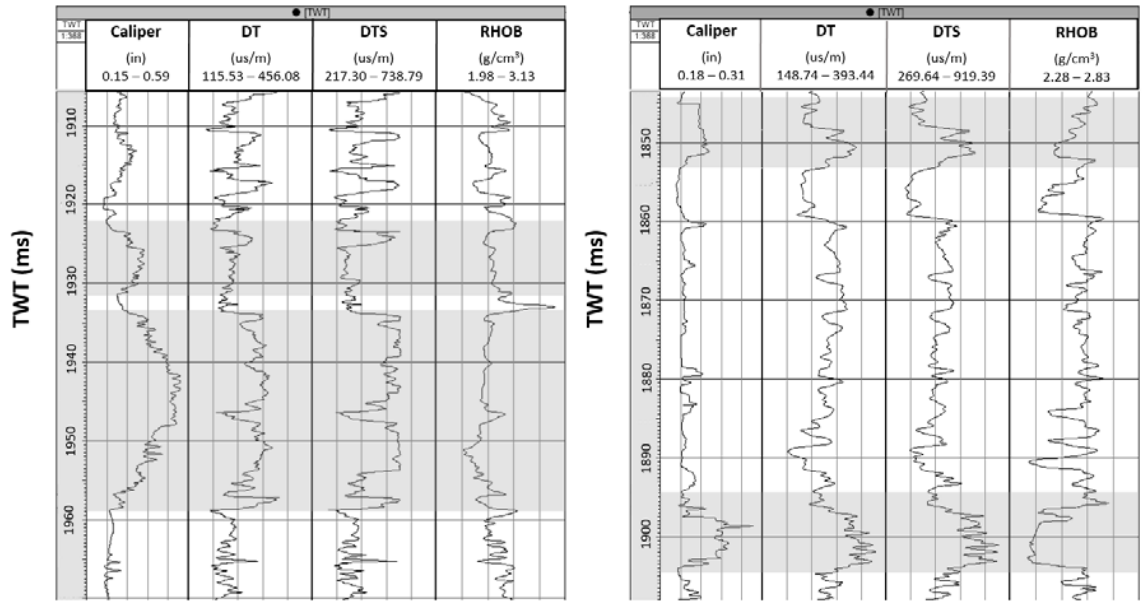

**Figure 2:** Two wells showing from the left to the right track: caliper, P-sonic, S-sonic and density logs. Gray background shadings highlight potential suspect log samples associated with collapsed sections as interpreted from the caliper log.

**Pereira et al** –

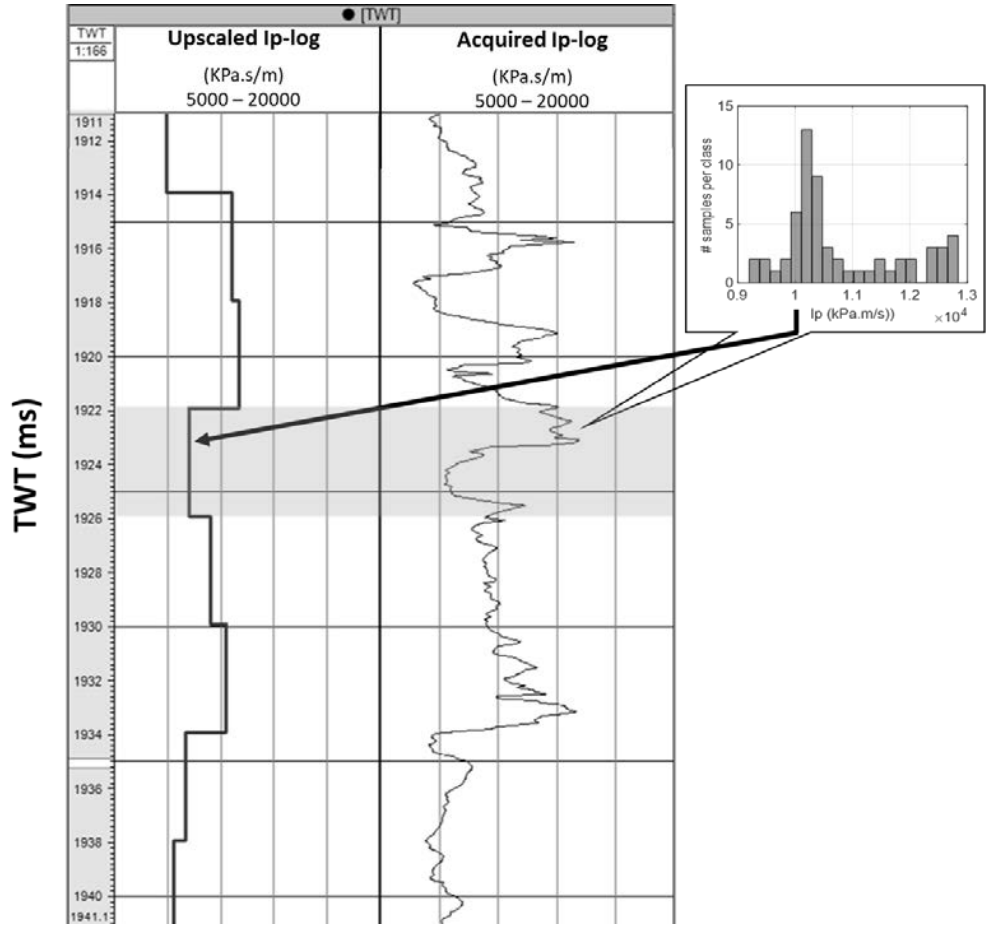

**Figure 3:** Comparison between upscaled (left track) and high-resolution Ip well-log data (right track). Gray background shading highlights the contrast of both logs at a given vertical cell of simulation grid.

**Pereira et al** –

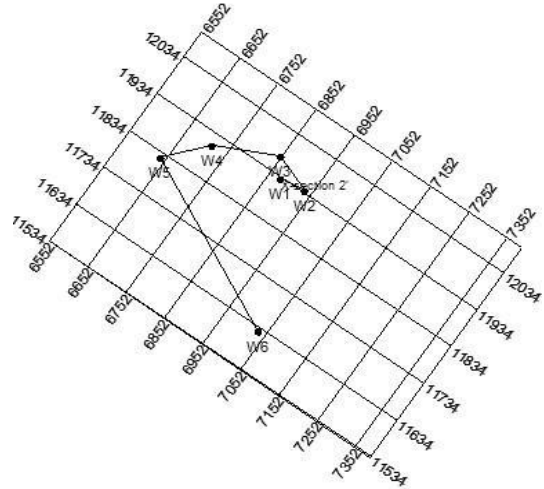

**Figure 4:** Geometry of the inversion grid showing the location of existing wells and a vertical well section shown to illustrate the results.

**Pereira et al** –

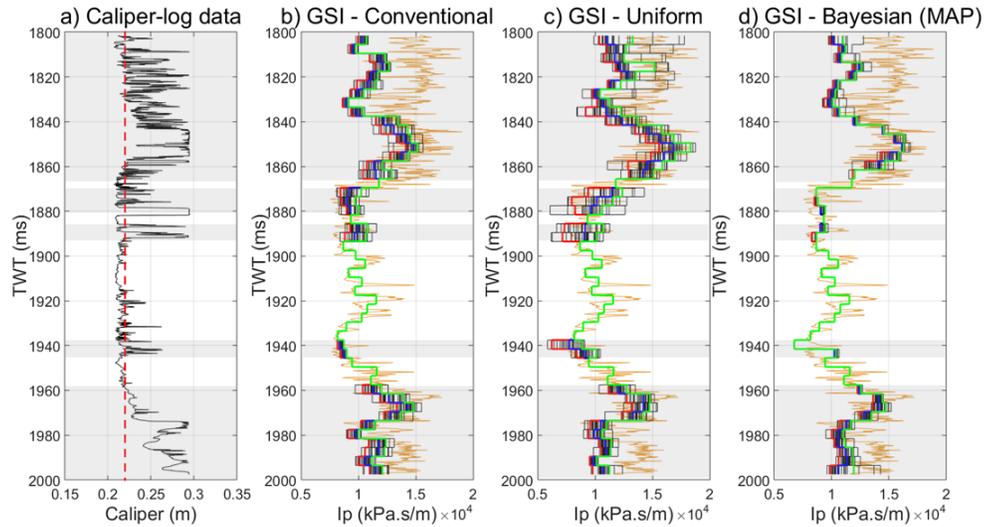

**Figure 5:** Well W4: high-resolution Ip-log is plotted in orange; 32 simulations in thin black; and P10, P50 and P90 percentiles in red, blue and green, respectively. Gray background shadings represents the location of suspect log samples. a) Caliper-log (black line) and threshold of 0.22m used to classify suspect log samples (red dashed line); Comparison between high-resolution log and inverted traces from: b) conventional; c) GSI with local uniform distributions; and d) GSI with local distributions provided by linearized Bayesian inversion.

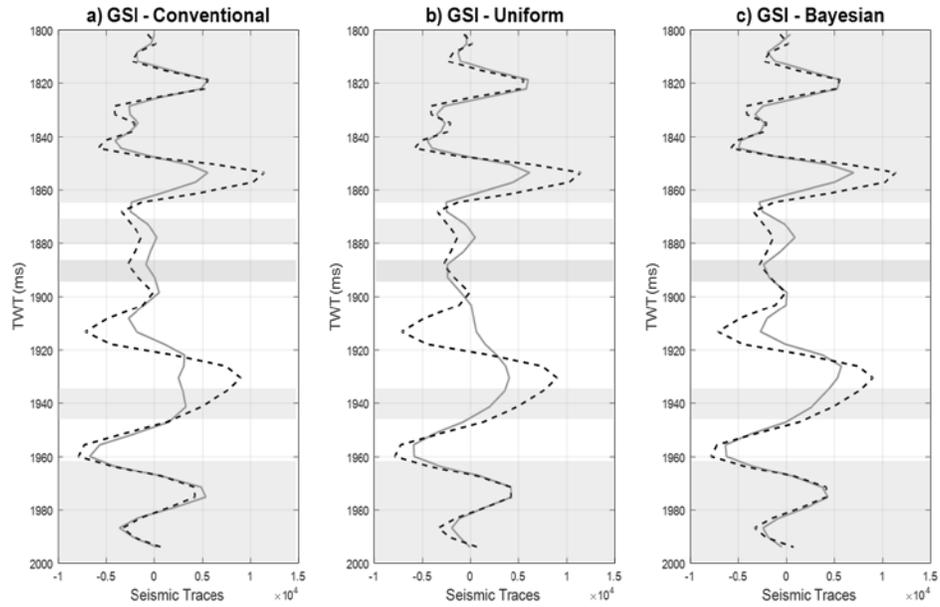

**Figure 6:** Well W4: gray background shadings represent the location of suspect log samples. Comparison between real seismic (black dashed line) and synthetic traces (gray line) computed from the average Ip trace generated during the last iteration of the inversion for: a) conventional GSI; b) GSI with local uniform distributions; and c) GSI with local distributions provided by linearized Bayesian inversion.

**Pereira et al** –

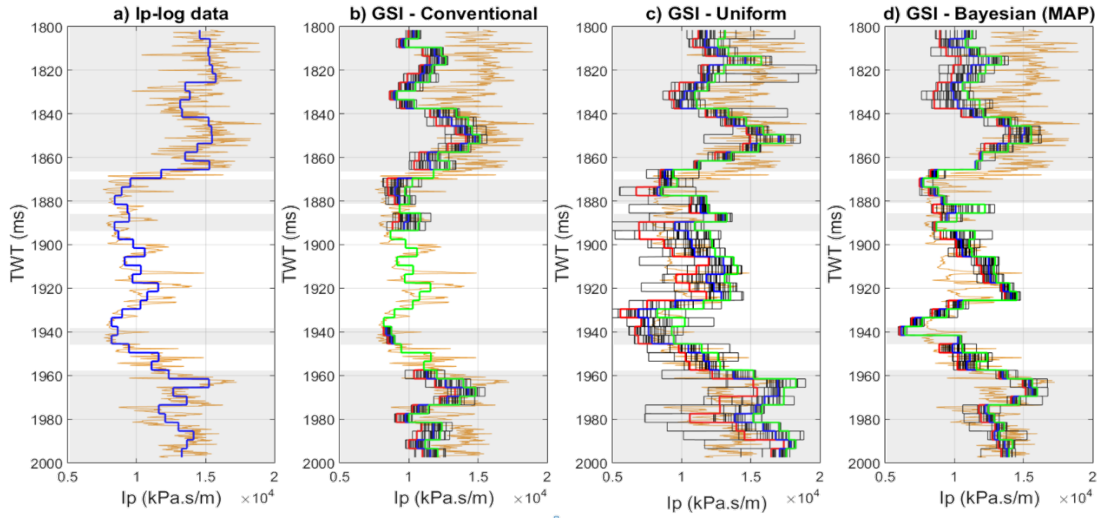

**Figure 7:** Well W4: gray background shadings represent suspect log samples that were removed; 32 simulations are plotted in thin black and P10, P50 and P90 percentiles in red, blue and green, respectively. a) High-resolution (in orange) and upscaled Ip-log data using Backus's averaging (in blue); Comparison between high-resolution log and inverted traces from: b) conventional GSI; c) GSI with local uniform distributions; and d) GSI with local distributions provided by linearized Bayesian inversion.

**Pereira et al** –

**Table 1:** Limitations criteria of the local correlation coefficient, from iteration-to-iteration, for the three scenarios during the seismic inversion process.

| Iterations | CC Limit 1 (%) | CC Limit 2 (%) | CC Limit 3(%) |
|:---:|:---:|:---:|:---:|
| 1 | 65.0 | 50.0 | 75.0 |
| 2 | 70.0 | 60.0 | 80.0 |
| 3 | 75.0 | 70.0 | 85.0 |
| 4 | 80.0 | 85.0 | 90.0 |
| 5 | 85.0 | 90.0 | 95.0 |
| 6 | 90.0 | 95.0 | 95.0 |

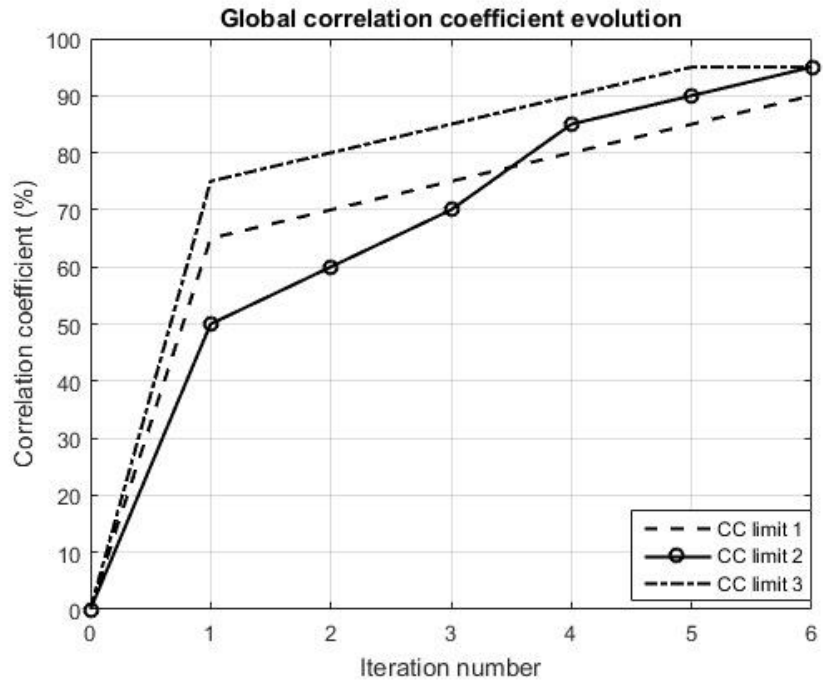

**Figure 8:** Evolution of the global CC between real and synthetic seismic volumes for the three scenarios shown in Table 1.

**Pereira et al** –

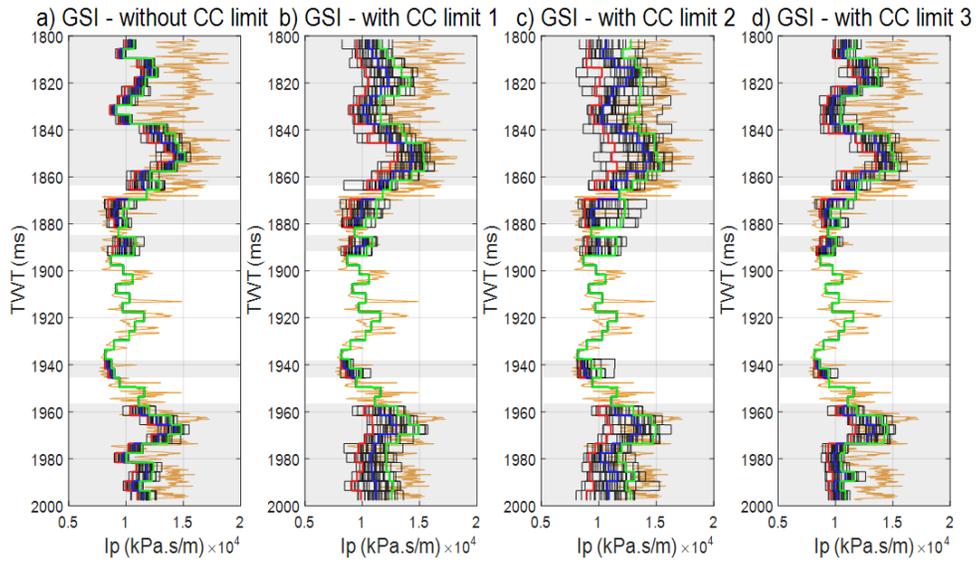

**Figure 9:** Well W4: gray background shadings represent suspect log samples; high-resolution Ip-log data is plotted in orange; 32 simulations in thin black; and P10, P50 and P90 percentiles in red, blue and green, respectively. Comparison between Ip traces inverted under the scenarios described in Table 1 for: a) conventional GSI without limitation criteria in the local CC; b) limit 1, c) limit 2 and d) limit 3.

**Pereira et al** –

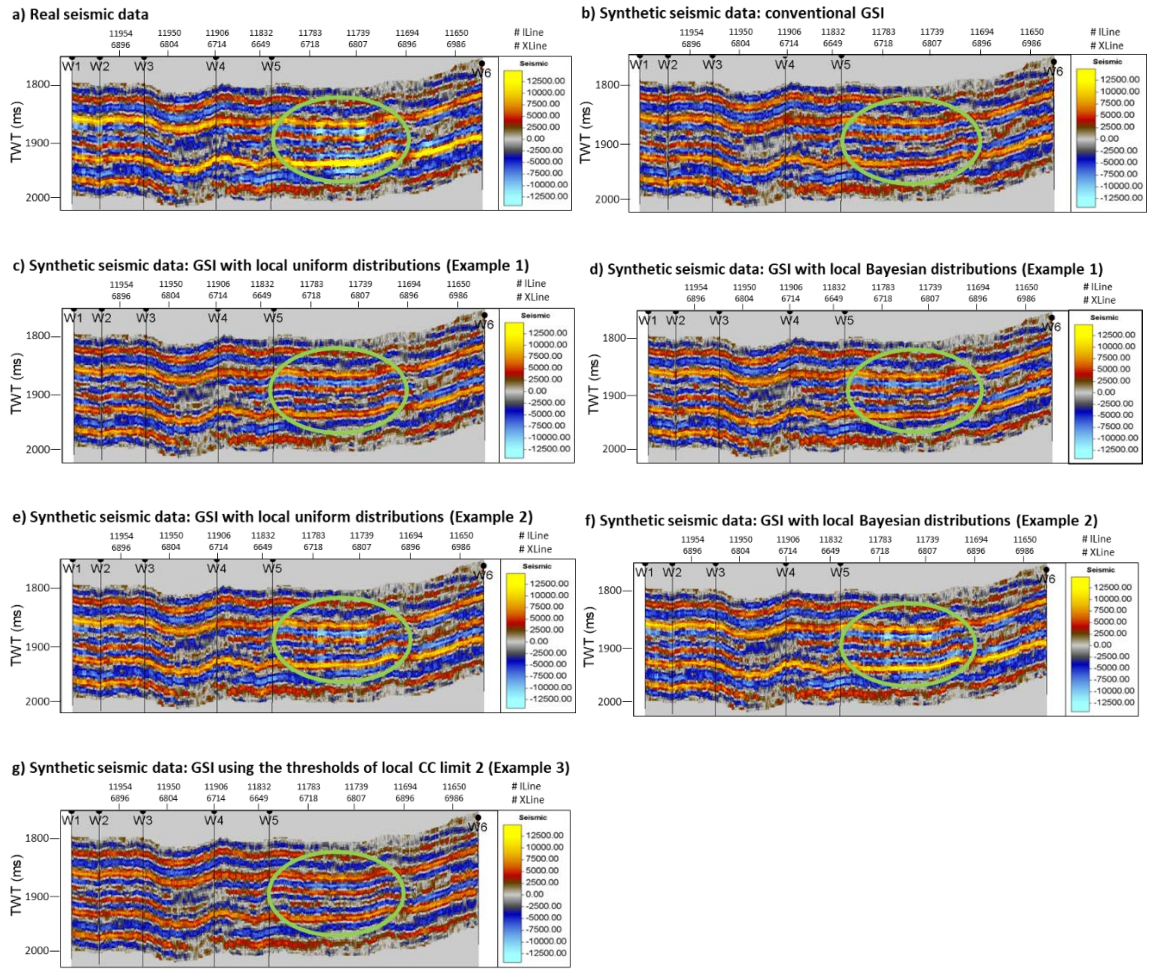

**Figure 10:** Vertical seismic sections of a) real seismic data and the synthetics of: b) conventional GSI; c) GSI with local uniform distributions and d) GSI with local distributions provided by linearized Bayesian (Example 1); e) GSI with local uniform distributions and f) GSI with local distributions computed from linearized Bayesian inversion (Example 2) and g) GSI using as cap the local CC limit 2 (Table 1) (Example 3).

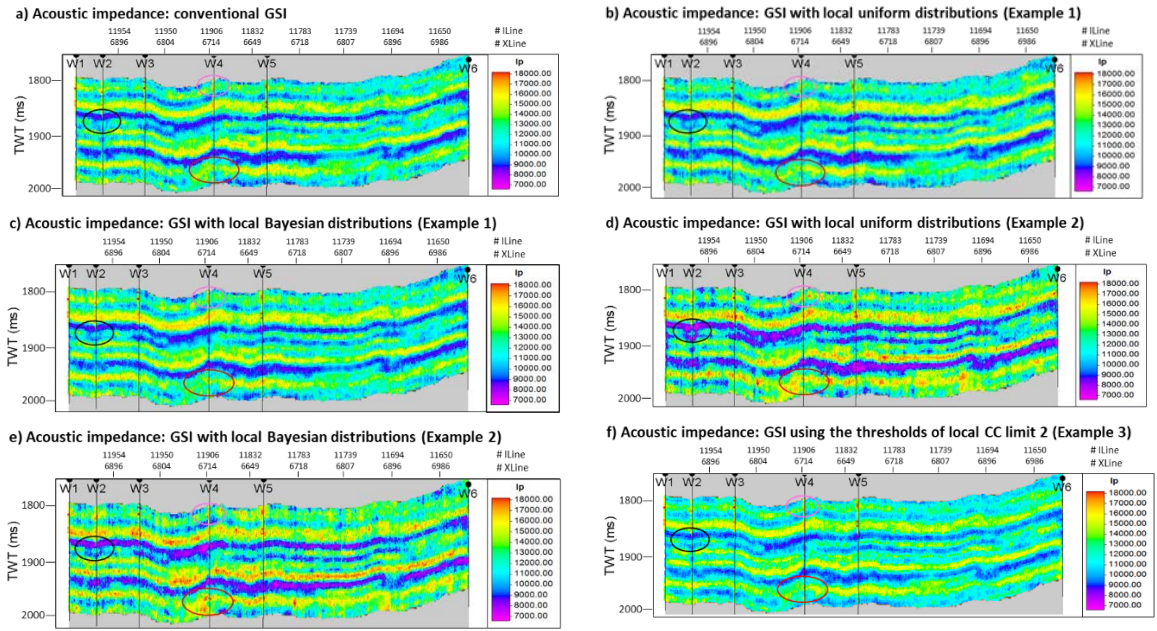

**Figure 11:** Vertical sections extracted from the average Ip model generated at the last iteration using: a) conventional GSI; b) GSI with local uniform distributions and c) GSI with local distributions provided by linearized Bayesian (Example 1); d) GSI with local uniform distributions and e) GSI with local distributions computed from linearized Bayesian inversion (Example 2) and f) GSI using as cap the local CC limit 2 (Table 1) (Example 3).

**Pereira et al** –

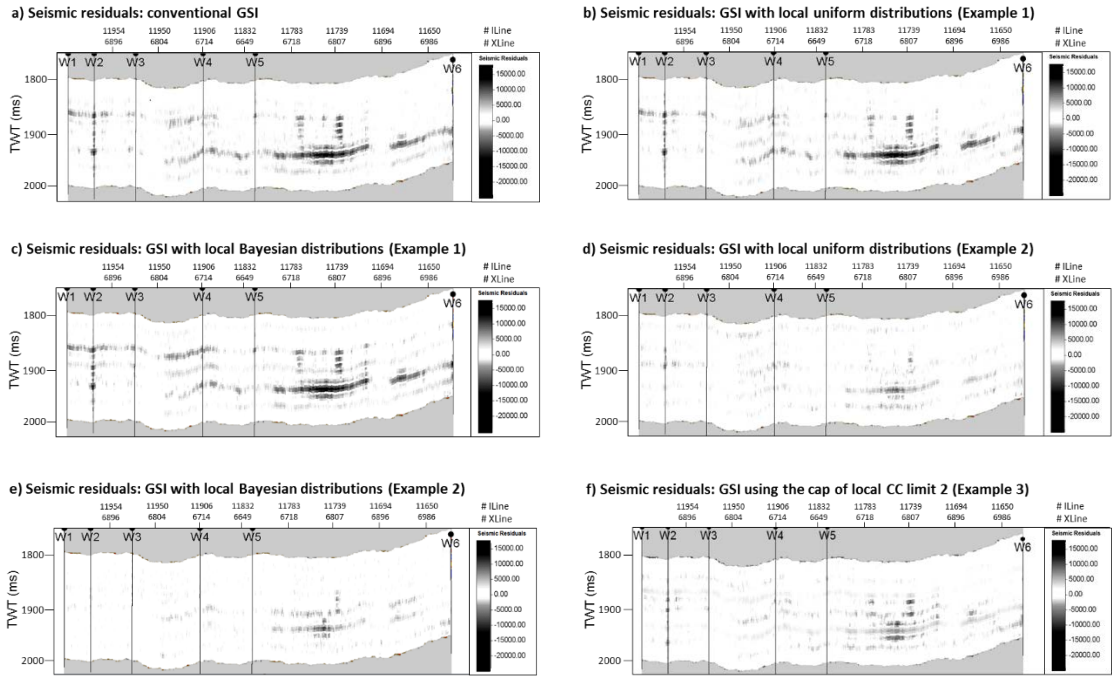

**Figure 12:** Vertical sections of amplitude residuals between real seismic data and synthetics of: a) conventional GSI; b) GSI with local uniform distributions and c) GSI with local distributions provided by linearized Bayesian (Example 1); d) GSI with local uniform distributions and e) GSI with local distributions computed from linearized Bayesian inversion (Example 2) )and f) GSI using as cap the local CC limit 2 (Table 1) (Example 3).

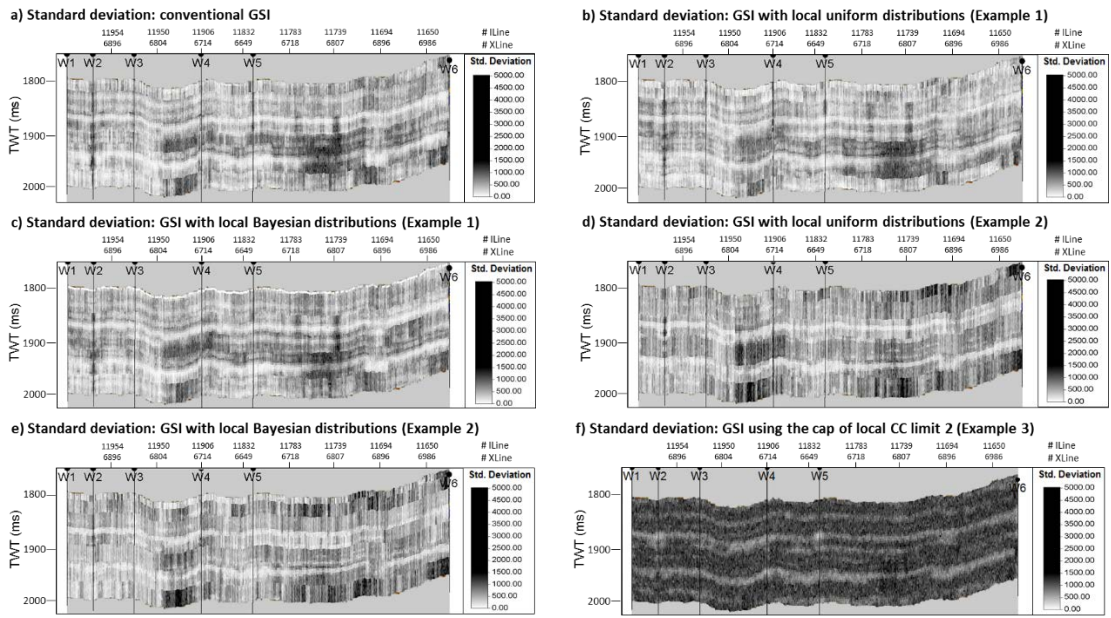

**Figure 13:** Vertical sections of standard deviation illustrating the variability of the ensemble of 32 Ip models generated at the last iteration using: a) conventional GSI; b) GSI with local uniform distributions and c) GSI with local distributions provided by linearized Bayesian (Example 1); d) GSI with local uniform distributions and e) GSI with local distributions computed from linearized Bayesian inversion (Example 2) and f) GSI using as cap the local CC limit 2 (Table 1) (Example 3).